\documentclass[12pt,a4paper,english,british]{article}
\usepackage[latin9]{inputenc}
\usepackage{mathrsfs}
\usepackage{bm}
\usepackage{amsmath}
\usepackage{amssymb}
\usepackage{graphicx}
\usepackage{esint}

\makeatletter

\newcommand*\LyXZeroWidthSpace{\hspace{0pt}}
\pdfpageheight\paperheight
\pdfpagewidth\paperwidth

\DeclareTextSymbolDefault{\textquotedbl}{T1}

\newcommand{\lyxaddress}[1]{
	\par {\raggedright #1
	\vspace{1.4em}
	\noindent\par}
}

\usepackage{latexsym}

\usepackage{amsmath}
\usepackage{amssymb}
\usepackage{booktabs}
\usepackage{multirow}
\usepackage{float} 
\usepackage{array} 
\usepackage{multicol} 

\usepackage{babel}

\makeatother

\usepackage{babel}
\begin{document}
\title{\selectlanguage{english}%
\begin{flushright}  {\small UWThPh 2025-11\,}  \bigskip \bigskip \bigskip \end{flushright}\foreignlanguage{british}{Exploring
Entanglement Entropy for a Particle on a Torus with constant Metric
and constant $U(1)$ - Gauge Field}\\
\foreignlanguage{british}{ }}
\author{Helmuth Huffel$^{\,\,1}$\thanks{helmuth.hueffel@univie.ac.at} ~and
Gerald Kelnhofer$^{\,\,1}$ }
\maketitle

\lyxaddress{\begin{center}
$^{1}$$\,$ {\small{}Faculty of Physics, University of Vienna,
}{\small Boltzmanngasse}{\small{} 5, A-1090 Vienna, Austria }\smallskip{}
 \\
 
\par\end{center}}
\begin{abstract}
\noindent In this paper, we focus on the entanglement entropy associated
with a particle confined to a torus with constant metric and $\theta$-terms
related to a constant external $U(1)$ - gauge field. Through this
investigation, we aim to elucidate the interplay between geometric
properties, topological terms, and entanglement entropy. Remarkably,
we find that the presence of non-vanishing entanglement entropy is
closely tied to fixing the metric. This insight highlights the significant
influence of quantum entanglement on our understanding of quantum
dynamics in complex spatial configurations. 
\end{abstract}

\section{Introduction}

Entanglement entropy (for recent reviews and exhaustive lists of references
see \cite{introduction 1,introduction 2,introduction 3}) is a pivotal
concept in quantum systems, providing insights into quantum correlations,
topological properties, and phase transitions. In quantum field theory
and many-body physics, it serves as a nonlocal measure of entanglement
between subsystems, and its application to systems with nontrivial
geometries or external gauge fields offers a unique opportunity to
explore how these factors influence quantum properties.\\
\\
Entanglement entropy establishes a fundamental connection to system
parameters. For instance, its well-documented relation with the central
charge in 1+1 dimensional conformal field theory \cite{Calabrese 1}
shows how critical properties are encoded in the entropy. Its extension
to finite-size systems, boundaries, and multiple disjoint intervals
\cite{Calabrese-2} further reveals a geometric dependence, and the
holographic approach based on the AdS/CFT correspondence \cite{Ryu}
links entanglement entropy to minimal surface areas in anti--de Sitter
space, thereby connecting it to spacetime geometry. Other notable
examples include quantum spin chains at criticality \cite{vidal}---where
the logarithmic scaling of entanglement entropy is governed by the
central charge and fermionic lattice models that obey an area law
with corrections reflecting the geometry of the Fermi surface \cite{eisert}.
These examples illustrate that entanglement entropy reflects system
parameters and captures essential features of the underlying quantum
state. \\
\\
Entanglement entropy is also often related to the presence of ground
state degeneracies, reflecting the underlying quantum structure. For
example, \cite{ares} explored the entanglement between the Maxwell
field and a non-relativistic charged point-particle, discussing entanglement
linked to ground state multiplicities; \cite{okatev} examined entropy
in Heisenberg ferromagnets under degeneracy-induced quantum criticality;
and \cite{hines} studied spin models, revealing changes in entanglement
near degeneracies. These insights parallel our study, where degeneracy-driven
features are shaped by entanglement entropy.\\
\\
Our work explores a system that combines geometric and gauge field
elements to elucidate how entanglement entropy reflects and constrains
a system\textquoteright s properties. We focus on a non-relativistic
charged particle coupled to an external constant U(1) gauge field
and confined to a 2-torus with a fixed metric. This setup provides
a platform to investigate how entanglement entropy depends on spatial
configurations and constrains model parameters.\\
\\
Under certain conditions of the external gauge field and the metric,
our model exhibits degenerate ground states---superpositions of momentum
eigenstates---that give rise to a non-vanishing entanglement entropy.
In this respect, our model provides an example of single-particle
entanglement, i.e., the quantum correlation between a particle's degrees
of freedom (here, its momentum components) {[}3{]}.\\
\\
By analysing the relationship between entanglement entropy and ground
state degeneracy, we investigate the system's parameter space$-$defined
by the U(1) gauge field $\bm{\theta}$ and the metric parameter $\lambda$.
Our findings reveal that entanglement entropy becomes nontrivial only
under specific parameter constraints: for twofold degeneracy the entropy
is nonzero in certain regions (with $\lambda$ varying within finite
bounds), for threefold degeneracy $\lambda$ is uniquely fixed for
given values of $\bm{\theta}$, and in the fourfold case both $\bm{\theta}$
and $\lambda$ are uniquely constrained with the metric becoming the
canonical metric on the 2-torus. \\
\\
These results underscore the interplay between entanglement entropy,
topology, and geometry, advancing our understanding of how quantum
correlations interrelate with geometric and topological principles
and providing a foundation for exploring the interdependence of these
features in quantum systems.

\section{The model}

We consider the quantum dynamics of a non-relativistic charged point
particle on a 2-torus, $\mathbb{T}^{2}=\mathbb{S}^{1}\times\mathbb{S}^{1}$,
under the influence of an external constant $U(1)$ - gauge field
$A=\sum_{j=1}^{2}\theta_{j}dx^{j}$, where $\bm{\theta=}\left(\theta_{1},\theta_{2}\right)\ensuremath{\in\mathrm{\mathcal{\mathscr{\mathbb{R}}}^{2}}}$.
Here $\mathrm{\boldsymbol{x}}=(x_{1,}x_{2})\in\mathbb{R}^{2}$ denotes
the coordinates on $\mathbb{T}^{2}$ with the identification $\mathrm{\boldsymbol{x}}\thicksim\mathrm{\boldsymbol{x}}+\mathrm{\boldsymbol{l}}$,
where $\boldsymbol{l}=(l_{1},l_{2})\in\mathbb{Z}^{2}$. The torus
is equipped with a constant metric $g=(g_{jk})$, given by

\begin{equation}
g=\begin{pmatrix}\frac{L_{1}^{2}}{q_{1}^{2}} & \frac{L_{1}L_{2}\lambda}{2}\\
\frac{L_{1}L_{2}\lambda}{2} & \frac{L_{2}^{2}}{q_{2}^{2}}
\end{pmatrix}.
\end{equation}
Here, $L_{1},\,L_{2}$ denote the lengths of the circles in $\mathbb{T}^{2}$,
and $q_{1},q_{2},\lambda\in\mathbb{R}$ are fixed parameters satisfying
$q_{1}>0,\:q_{2}>0$ and $\varLambda:=\frac{1}{q_{1}^{2}q_{2}^{2}}-\frac{\lambda^{2}}{4}>0$.
The volume of $\mathbb{T}^{2}$ with respect to the metric is $Vol(\mathbb{T}^{2})=L_{1}L_{2}\varLambda^{\frac{1}{2}}$.\\
\\
The Lagrangian of the model is given by

\begin{equation}
L\left(\mathrm{\boldsymbol{x}},\dot{\mathrm{\boldsymbol{x}}}\right)=\frac{1}{2}\ \sum_{j,k=1}^{2}g_{jk}\dot{x}_{j}\dot{x}_{k}+\text{\ensuremath{\theta_{j}}}\dot{x}_{j}.\label{eq:lagrange}
\end{equation}
This system can also be interpreted as a model of two interacting
particles on a circle with $\theta$-terms, where the coupling strength
is determined by $\lambda$. \\
\\
The canonical momentum $\mathbf{p}=\left(p_{1},p_{2}\right)\in\mathbb{R}^{2}$
conjugate to $\mathrm{\boldsymbol{x}}$ is 
\begin{equation}
p_{j}=\sum_{k=1}^{2}g_{jk}\dot{x}_{k}+\text{\ensuremath{\theta_{j}}}
\end{equation}
and the classical Hamiltonian becomes
\begin{equation}
\begin{split}H\left(\mathrm{\boldsymbol{x}},\mathbf{p}\right)= & \frac{1}{2}\,\sum_{j,k=1}^{2}g_{jk}^{-1}\left(p_{j}-\theta_{j}\right)\left(p_{k}-\theta_{k}\right)\\
= & \frac{1}{2\varLambda L_{1}L_{2}}\Big(\frac{L_{2}}{L_{1}}\frac{1}{q_{2}^{2}}\left(p_{1}-\theta_{1}\right)^{2}+\frac{L_{1}}{L_{2}}\frac{1}{q_{1}^{2}}\left(p_{2}-\theta_{2}\right)^{2}-\lambda\left(p_{1}-\theta_{1}\right)\left(p_{2}-\theta_{2}\right)\Big).
\end{split}
\end{equation}
The classical equations of motion $\dot{p}_{j}=0,j=1,2$, indicate
that the particle moves freely on the torus. Here, the two momentum
components remain independent, with no inherent interdependence. Our
goal is to examine whether this persists in the corresponding quantum
theory. To answer this question, we specifically analyse how the quantised
system responds to the external parameters $\bm{\theta}$ and $\lambda.$
\\
\\
In the quantised model the space of states is the Hilbert space $\mathfrak{h}=L^{2}(\mathbb{T}^{2},\mathbb{C})$,
the space of square-integrable complex-valued functions on the torus
$\mathbb{T}^{2}$, represented by functions $\psi:\mathrm{\mathcal{\mathscr{\mathbb{R}}}^{2}}\rightarrow\mathbb{C},$
such that
\begin{equation}
\psi(\boldsymbol{x}+\boldsymbol{l})=\psi(\mathbf{\boldsymbol{x}}),\enspace\forall\,\boldsymbol{l}\in\mathbb{Z}^{2}.
\end{equation}
 The inner product on $\mathfrak{h}$ is defined by
\begin{equation}
\langle\phi|\chi\rangle=\intop_{0}^{1}d\mathrm{\boldsymbol{x}}\left(det\,g\right)^{\frac{1}{2}}\phi_{\boldsymbol{}}^{*}(\mathrm{\boldsymbol{x}})\,\chi_{\boldsymbol{}}(\mathrm{\boldsymbol{x}})
\end{equation}
for $\phi,\chi\in\mathfrak{h}.$ In the Schr\"odinger representation,
the canonical momentum operators, defined by 
\begin{equation}
\hat{p}_{j}^{\bm{\theta}}=\frac{1}{i}\nabla_{j}^{\bm{\theta}},\quad\quad\nabla_{j}^{\bm{\theta}}=\frac{\partial}{\partial x_{j}}-i\theta_{j},\quad\quad j=1,2,
\end{equation}
are self-adjoint operators on $\mathfrak{h},$ where $\nabla_{j}^{\bm{\theta}}$
is the covariant derivative associated with the flat gauge field $A.$
The quantum Hamiltonian on $\mathfrak{h}$ is the operator $\hat{H}^{(\bm{\theta},\lambda)}$

\begin{equation}
\hat{H}^{(\bm{\theta},\lambda)}=-\frac{1}{2}\ \sum_{j,k=1}^{2}g_{jk}^{-1}\nabla_{j}^{\bm{\theta}}\nabla_{k}^{\bm{\theta}},\label{eq:hamiltonian}
\end{equation}
where we have explicitly indicated the parameters. \\
\\
Notice that there exists an alternative formulation of the model that
does not rely on the presence of $\bm{\theta}$-terms in the Lagrangian.
Consider the Hilbert space $\mathfrak{h_{\bm{\theta}}}$ of functions
$\chi:\mathrm{\mathcal{\mathscr{\mathbb{R}}}^{2}}\rightarrow\mathbb{C}$
satisfying the quasi-periodicity condition 
\begin{equation}
\chi(\boldsymbol{x}+\boldsymbol{l})=e^{-i\bm{\theta}.\boldsymbol{l}}\chi(\mathbf{\boldsymbol{x}}),\enspace\forall\,\boldsymbol{l}\in\mathbb{Z}^{2}.
\end{equation}
 There exists a unitary operator $U_{\bm{\theta}}:\mathfrak{h}\rightarrow\mathfrak{h_{\bm{\theta}}}$
defined by
\begin{equation}
\left(U_{\bm{\theta}}\psi\right)(\mathrm{\boldsymbol{x}}):=e^{-i\bm{\theta}.\boldsymbol{\mathbf{\boldsymbol{x}}}}\psi(\mathbf{\boldsymbol{x}}).
\end{equation}
Since $U_{\bm{\theta}}\,\nabla_{j}^{\bm{\theta}}\,U_{\bm{\theta}}^{-1}=\frac{\partial}{\partial x_{j}}$
on $\mathfrak{h_{\bm{\theta}}},$ the $\bm{\theta}$-dependent terms
in the Hamiltonian $\hat{H}^{(\bm{\theta},\lambda)}$ can be eliminated
using this transformation, yielding a new Hamiltonian $\hat{H}^{(\lambda)}$
\begin{equation}
\hat{H}^{(\lambda)}:=U_{\bm{\theta}}\,\hat{H}^{(\bm{\theta},\lambda)}\,U_{\bm{\theta}}^{-1}=-\frac{1}{2}\ \sum_{j,k=1}^{2}g_{jk}^{-1}\frac{\partial}{\partial x_{j}}\frac{\partial}{\partial x_{k}}.
\end{equation}
Thus the physical systems described by $\hat{H}^{(\bm{\theta},\lambda)}$
and $\hat{H}^{(\lambda)}$ are equivalent. In the remainder of this
paper, we continue using the formulation involving $\hat{H}^{(\bm{\theta},\lambda)}$.
\\
\\
The eigenstates of the momentum operator $\hat{p}_{j}^{\bm{\theta}}$
and the Hamiltonian $\hat{H}^{(\bm{\theta},\lambda)}$ are labelled
by $\boldsymbol{m}=(m_{1},m_{2})\in\mathbb{Z}^{2}$ and denoted by
$\vert\boldsymbol{m}\rangle$. In the coordinate representation the
normalised eigenstates $\psi_{\boldsymbol{m}}\in\mathfrak{h}$ take
the form
\begin{equation}
\psi_{\boldsymbol{m}}(x_{1,}x_{2})\equiv\psi_{\boldsymbol{m}}(\mathrm{\boldsymbol{x}})=\langle\mathrm{\boldsymbol{x}}|\mathrm{\boldsymbol{m}}\rangle=\left(L_{1}L_{2}\right)^{-\frac{1}{2}}\varLambda^{-\frac{1}{4}}\,e^{2\pi i\boldsymbol{m}.\mathrm{\boldsymbol{x}}}.
\end{equation}
Evidently, the eigenstates factorise into the tensor product
\begin{equation}
\vert\boldsymbol{m}\rangle=\vert m_{1}\rangle^{(1)}\otimes\vert m_{2}\rangle^{(2)}.
\end{equation}
 The eigenvalues of $\hat{H}^{(\bm{\theta},\lambda)}$ are obtained
as
\begin{equation}
E_{\boldsymbol{m}}\left(\bm{\theta},\lambda\right)=\frac{2\pi^{2}}{\varLambda L_{1}L_{2}}\Big(\frac{L_{2}}{L_{1}}\frac{1}{q_{2}^{2}}(\text{\ensuremath{m_{1}}}-\frac{\text{\ensuremath{\theta_{1}}}}{2\pi})^{2}+\frac{L_{1}}{L_{2}}\frac{1}{q_{1}^{2}}(\text{\ensuremath{m_{2}}}-\frac{\text{\ensuremath{\theta_{2}}}}{2\pi})^{2}-\lambda(m_{1}-\frac{\text{\ensuremath{\theta_{1}}}}{2\pi})(m_{2}-\frac{\text{\ensuremath{\theta_{2}}}}{2\pi})\Big)
\end{equation}
The spectrum exhibits a symmetry under shifts of $\theta_{j}$ by
multiples of $2\pi$, 
\begin{equation}
E_{\boldsymbol{m}}\left(\bm{\theta}+2\pi\,\bm{l},\lambda\right)=E_{\bm{m}-\bm{l}}\left(\bm{\theta},\lambda\right),\qquad\forall\,\boldsymbol{l}\in\mathbb{Z}^{2}.\label{eq:shift}
\end{equation}
Let us define the unitary operator $\left(U_{\bm{l}}\psi\right)(\mathrm{\boldsymbol{x}})=e^{2\pi i\bm{l}\mathrm{\boldsymbol{x}}}\psi(\mathrm{\boldsymbol{x}})$
with $\psi\in\mathfrak{h}$ and $\boldsymbol{l}\in\mathbb{Z}^{2}.$
Then
\begin{equation}
U_{\boldsymbol{l}}^{-1}\hat{H}^{(\bm{\theta},\lambda)}U_{\boldsymbol{l}}=\hat{H}^{(\bm{\theta}+2\pi\bm{l},\lambda)}
\end{equation}
shows that the quantum theories of the classical system (\ref{eq:lagrange})
associated with $\bm{\theta}$ and $\bm{\theta}+2\pi\boldsymbol{l}$
are unitarily equivalent. The inequivalent theories are thus classified
by $\bm{\theta}\:mod\,2\pi\,\mathbb{Z}^{2},$ while also ensuring
that $\lambda$ satisfies $det\,g>0.$ Hence the parameter space $\mathcal{\mathfrak{R}}$
of the family of quantum systems is 
\begin{equation}
\mathcal{\mathfrak{R}}=\Big\{(\bm{\theta},\lambda)\,|\,\bm{\theta\in}\mathbb{T}^{2},\;det\,g>0\Big\}.\label{eq:rg=00005Censuremath=00007B=00005Ctheta=00007D}
\end{equation}
Later, we will consider the special case $L_{1}=L_{2}=q_{1}=q_{2}=1$.
Then, with an appropriate parametrisation for $\bm{\theta}$ we have
for the parameter space \\
\begin{equation}
\mathcal{\mathfrak{R}}=\Big\{(\bm{\theta},\lambda)\,|\,-\pi\leq\text{\ensuremath{\theta_{1}}}<\pi,-\pi\leq\text{\ensuremath{\theta_{2}}}<\pi,-2<\lambda<2\Big\}.\label{eq:rg=00005Censuremath=00007B=00005Ctheta=00007D-1}
\end{equation}
To provide a comprehensive discussion of the model we diagonalise
the Hamiltonian by performing a symplectic transformation of the original
variables $\mathrm{\boldsymbol{x}}$ and $\mathbf{p}$ to adapted
variables $\mathbf{\tilde{\mathrm{\boldsymbol{x}}}}=\left(\tilde{x}_{1},\tilde{x}_{2}\right)$
and $\tilde{\mathrm{\boldsymbol{p}}}=\left(\tilde{p}_{1},\tilde{p}_{2}\right)$
\begin{equation}
\mathbf{\tilde{\mathrm{\boldsymbol{x}}}}=\sigma\mathrm{\boldsymbol{x}}\,\in\mathbb{R}^{2}\;\textrm{and}\;\tilde{\mathrm{\boldsymbol{p}}}=\left(\sigma^{-1}\right)^{T}\mathbf{p}\,\in\mathbb{R}^{2},
\end{equation}
 where 
\begin{equation}
\sigma=\frac{1}{\sqrt{2}}\left(\begin{array}{cc}
\sqrt{\frac{L_{1}q_{2}}{L_{2}q_{1}}} & \sqrt{\frac{L_{2}q_{1}}{L_{1}q_{2}}}\\
-\sqrt{\frac{L_{1}q_{2}}{L_{2}q_{1}}} & \sqrt{\frac{L_{2}q_{1}}{L_{1}q_{2}}}
\end{array}\right).
\end{equation}
We have 

\begin{equation}
\sigma^{T}J\sigma=J,\qquad\textrm{for}\quad J=\left(\begin{array}{cc}
0 & 1\\
-1 & 0
\end{array}\right)
\end{equation}
and indeed it holds that 
\begin{equation}
\sigma\,g^{-1}\sigma^{T}=\frac{1}{L_{1}L_{2}}\left(\begin{array}{cc}
\frac{1}{\tilde{\Lambda}_{1}} & 0\\
0 & \frac{1}{\tilde{\Lambda}_{2}}
\end{array}\right)
\end{equation}
where
\begin{equation}
\tilde{\Lambda}_{1}=\frac{1}{q_{1}q_{2}}+\frac{\lambda}{2},\qquad\tilde{\Lambda}_{2}=\frac{1}{q_{1}q_{2}}-\frac{\lambda}{2}.
\end{equation}
For notational convenience let us introduce adapted angles $\mathbf{\mathbf{\mathbf{\tilde{\bm{\theta}}}}}=\left(\tilde{\theta}_{1},\tilde{\theta}_{2}\right)$
\begin{equation}
\mathbf{\mathbf{\mathbf{\tilde{\bm{\theta}}}}}=\frac{1}{\sqrt{2}\pi}\left(\sigma^{-1}\right)^{T}\,\bm{\theta}\,\in\mathbb{R}^{2},
\end{equation}
then the Hamiltonian in adapted coordinates $\tilde{H}=\left(\sigma^{-1}\times\sigma^{T}\right)^{*}\,H$
becomes

\begin{equation}
\tilde{H}\left(\tilde{\mathrm{\boldsymbol{x}}},\tilde{\mathrm{\boldsymbol{p}}},\mathbf{\mathbf{\mathbf{\mathbf{\tilde{\bm{\theta}}}}}},\lambda\right)=\frac{1}{2L_{1}L_{2}}\left(\frac{1}{\tilde{\Lambda}_{1}}\left(\tilde{p}_{1}-\sqrt{2}\pi\tilde{\theta}_{1}\right)^{2}+\frac{1}{\tilde{\Lambda}_{2}}\left(\tilde{p}_{2}-\sqrt{2}\pi\tilde{\theta}_{2}\right)^{2}\right)
\end{equation}
Given that $\mathbf{\tilde{\mathrm{\boldsymbol{x}}}}$ and $\ensuremath{\tilde{\mathrm{\boldsymbol{p}}}}$
are canonical variables, in the Schr\"odinger representation, $\tilde{p}_{k}$
transforms into $\frac{1}{i}\frac{\partial}{\partial\tilde{x}_{k}}$.
It is acting on wave functions $\phi_{\boldsymbol{\tilde{m}}}\left(\mathbf{\tilde{\mathrm{\boldsymbol{x}}}}\right)$,
which have to satisfy $\phi_{\boldsymbol{\tilde{m}}}(\mathbf{\tilde{\mathrm{\boldsymbol{x}}}}+\tilde{\mathrm{\boldsymbol{l}}})=\phi_{\boldsymbol{\tilde{m}}}(\mathbf{\tilde{\mathrm{\boldsymbol{x}}}})$.
Here $\tilde{\mathrm{\boldsymbol{l}}}=\left(\tilde{l}_{1},\tilde{l}_{2}\right)\in\mathscr{\tilde{L}}$,
where $\mathscr{\tilde{L}}$ now is the lattice
\begin{equation}
\mathscr{\tilde{L}}=\left\{ \tilde{\mathrm{\boldsymbol{l}}}=\sigma\boldsymbol{l}|\boldsymbol{l}\in\mathbb{Z}^{2}\right\} =\sigma\left(\mathscr{\mathbb{Z}}^{2}\right)
\end{equation}
We have the diffeomorphism $\tau:\;\mathscr{\mathbb{R}}^{2}/\mathscr{\mathbb{Z}}^{2}\longrightarrow\mathscr{\mathbb{R}}^{2}/$$\tilde{\mathscr{L}}$,
with $\tau\left(\left[\boldsymbol{x}\right]_{\mathscr{\mathbb{Z}}^{2}}\right)=\left[\sigma\mathrm{\boldsymbol{x}}\right]_{\mathscr{\tilde{L}}},$
as well as $\tau^{-1}\left(\left[\boldsymbol{\tilde{x}}\right]_{\mathscr{\mathscr{\tilde{L}}}}\right)=\left[\sigma^{-1}\mathrm{\boldsymbol{\tilde{x}}}\right]_{\mathscr{\mathbb{Z}}^{2}}$.
\\
\\
For simplification of notation, we also introduce adapted coordinates
$\boldsymbol{\tilde{m}}=\left(\tilde{m}_{1},\tilde{m}_{2}\right)$
\begin{equation}
\boldsymbol{\tilde{m}}=\sqrt{2}\,\left(\sigma^{-1}\right)^{T}\,\boldsymbol{m}\in\sqrt{2}\,\mathscr{\tilde{L}}^{*}
\end{equation}
via the mapping $\kappa:\:\mathscr{\mathbb{Z}}^{2}\longrightarrow$$\sqrt{2}\,\mathscr{\tilde{L}^{*}}$,
with $\kappa\left(\boldsymbol{m}\right)=\sqrt{2}\,\left(\sigma^{-1}\right)^{T}\,\boldsymbol{m}$.
Here $\mathscr{\tilde{L}}^{*}$ is the reciprocal lattice of $\mathscr{\tilde{L}}.$\\
\\
In adapted coordinates, the eigenstates $\ensuremath{\phi_{\boldsymbol{\tilde{m}}}(\mathbf{\tilde{\mathrm{\boldsymbol{x}}}})}$
are expressed with respect to the eigenstates $\psi_{\boldsymbol{m}}(\mathrm{\boldsymbol{x}})$
of the Hamiltonian $\hat{H}$ as\begin{equation}
\begin{aligned}\phi_{\boldsymbol{\tilde{m}}}\left(\mathbf{\tilde{\mathrm{\boldsymbol{x}}}}\right) & =\left(\tau^{-1\star}\psi_{\kappa^{-1}(\boldsymbol{\tilde{m}})}\right)(\mathrm{\boldsymbol{\tilde{x}}})=\psi_{\kappa^{-1}(\boldsymbol{\tilde{m}})}(\mathrm{\sigma^{-1}\mathrm{\boldsymbol{\tilde{x}}}})=\psi_{\frac{1}{\sqrt{2}}\sigma^{T}\boldsymbol{m}}(\mathrm{\sigma^{-1}\mathrm{\boldsymbol{\tilde{x}}}})\\
 & =\left(L_{1}L_{2}\right)^{-\frac{1}{2}}\varLambda^{-\frac{1}{4}}\,e^{\sqrt{2}\pi i\boldsymbol{\tilde{m}}.\mathbf{\tilde{\mathrm{\boldsymbol{x}}}}}.
\end{aligned}
\end{equation}Finally, we obtain the spectrum in diagonalised form
\begin{equation}
\tilde{E}_{\boldsymbol{\tilde{m}}}\left(\bm{\tilde{\theta}},\lambda\right)=\frac{\pi^{2}}{L_{1}L_{2}}\left(\frac{(\tilde{m}_{1}-\text{\ensuremath{\tilde{\theta}_{1}}})^{2}}{\tilde{\Lambda}_{1}}+\frac{(\tilde{m}_{2}-\text{\ensuremath{\tilde{\theta}_{2}}})^{2}}{\tilde{\Lambda}_{2}}\right)
\end{equation}
\\
In this paper, we will henceforth limit our analysis to the case where
$L_{1}=L_{2}=q_{1}=q_{2}=1,$ although the model with general parameters
$L_{1},L_{2},q_{1},q_{2}$ may offer interesting avenues for further
exploration. \\
\\
The restriction to $L_{1}=L_{2}=q_{1}=q_{2}=1$ is not merely a simplification
for computational convenience; rather, it allows for the solution
of the model in the sense that it enables a complete discussion of
the ground states and their degeneracies (see the next section and
specifically Lemmas 1--4 below).\\
\\
For later convenience, and using the notation 
\begin{equation}
E_{m_{1},m_{2}}(\bm{\theta},\lambda):=E_{\boldsymbol{m}}(\bm{\theta},\lambda),
\end{equation}
we recall the energy eigenvalues in the original coordinates with
simplified parameters:\begin{equation}
\begin{aligned}
E_{m_1, m_2}(\bm{\theta}, \lambda) &= \frac{1}{\pi^2} \Big( \theta_1^2 + 2\pi m_1 \left( \lambda \left( \theta_2 - 2\pi m_2 \right) - 2\theta_1 \right) \\
& \quad + \left( 2\pi m_2 - \theta_2 \right) \left( \theta_1 \lambda - \theta_2 + 2\pi m_2 \right) + 4\pi^2 m_1^2 \Big)
\end{aligned}
\end{equation} 

\section{Ground States and Degeneracies}

To systematically analyze the structure of the ground states and their
degeneracies, we present four key lemmas that establish the ordering
of energy levels. For brevity, the proofs are provided in the appendix.
Following these lemmas, we conclude with a theorem on nondegenerate
ground states.\subsection{Ordering of  Energy Levels}
\medskip{}
\begin{quote}
\textbf{Lemma 1}  \newline
For all \((\bm{\theta}, \lambda) \in \mathcal{\mathfrak{R}}\), and for all \(\boldsymbol{m} = (m_1, m_2) \in \mathbb{Z}^{2}\) with \(\mbox{\(\big||m_1| - |m_2|\big| > 1\)}\), one has   
\begin{equation}
E_{0,0}(\bm{\theta}, \lambda)<E_{m_1, m_2}(\bm{\theta}, \lambda).
\end{equation}
\end{quote}

\newpage

\begin{quote}
\textbf{Lemma 2} \begin{itemize}
 \item For all \((\bm{\theta}, \lambda) \in \mathcal{\mathfrak{R}}\) with \(\bm{\theta} \neq \bm{\bar{\theta}}=(-\pi,-\pi)\), and for all \(\boldsymbol{m} = (m_1, m_2) \in \mathbb{Z}^{2}\) with \(|m_1| - |m_2| =0\) and \(\boldsymbol{m} \neq (0,0)\),   one has
\begin{equation}
E_{0,0}(\bm{\theta}, \lambda)<E_{m_1, m_2}(\bm{\theta}, \lambda).
\end{equation}
\item
For all \(\lambda\) with $-2<\lambda<2$, and for all \(\boldsymbol{m} = (m_1, m_2) \in \mathbb{Z}^{2}\) with \(|m_1| - |m_2| =0\) and  \(\boldsymbol{m} \neq (0,0)\), \(\boldsymbol{m} \neq (-1,-1)\), one has
 \begin{equation}
 E_{0,0}(\bm{\bar{\theta}}, \lambda)=E_{-1,-1}(\bm{\bar{\theta}}, \lambda)<E_{m_1, m_2}(\bm{\bar{\theta}}, \lambda).
    \end{equation}
\end{itemize}
\end{quote}

\medskip{}

\begin{quote}
\textbf{Lemma 3} \newline 
For all $(\bm{\theta}, \lambda) \in \mathcal{\mathfrak{R}}$, and for all $\boldsymbol{m} = (m_1, m_2) \in \mathbb{Z}^2$ with
\begin{itemize}
    \item $\,|m_1| - |m_2|\, = 1$, $ m_1 > 0$, and  \(\boldsymbol{m} \neq (1,0)\), one has
    \begin{equation}
    E_{1,0}(\bm{\theta}, \lambda)<E_{m_1, m_2}(\bm{\theta}, \lambda).
    \end{equation}
    \item $\,|m_1| - |m_2|\, = 1$, $ m_1 < 0$, and  \(\boldsymbol{m} \neq (-1,0)\), one has
    \begin{equation}
 E_{-1,0}(\bm{\theta}, \lambda)<E_{m_1, m_2}(\bm{\theta}, \lambda).
    \end{equation}
\end{itemize}
\end{quote}

\medskip{}

\begin{quote}
\textbf{Lemma 4} \newline 
For all $(\bm{\theta}, \lambda) \in \mathcal{\mathfrak{R}}$, and for all $\boldsymbol{m} = (m_1, m_2) \in \mathbb{Z}^2$ with
\begin{itemize}
    \item $\,|m_1| - |m_2|\, = -1$,  $ m_2 > 0$, and  \(\boldsymbol{m} \neq (0,1)\), one has
    \begin{equation}
    E_{0,1}(\bm{\theta}, \lambda)<E_{m_1, m_2}(\bm{\theta}, \lambda).
    \end{equation}
    \item $\,|m_1| - |m_2|\, = -1$,  $ m_2 < 0$, and  \(\boldsymbol{m} \neq (0,-1)\), one has
    \begin{equation}
 E_{0,-1}(\bm{\theta}, \lambda)<E_{m_1, m_2}(\bm{\theta}, \lambda).
    \end{equation}
\end{itemize}
\end{quote}
\medskip{}We define the set 
\begin{equation}
\mathfrak{\mathcal{K}} = \{ \boldsymbol{k}^{(\alpha)} \mid \alpha = 0,1,2,\dots,5 \}
\end{equation}
of specific integer pairs that naturally arise from the above lemmas. In particular, we have

\begin{equation}
\begin{aligned}
\boldsymbol{k}^{(1)} &= (0,0), \\
\boldsymbol{k}^{(2)} &= (-1,0), \\
\boldsymbol{k}^{(3)} &= (1,0), \\
\boldsymbol{k}^{(4)} &= (0,-1), \\
\boldsymbol{k}^{(5)} &= (0,1),
\end{aligned}
\label{eq:K12345}
\end{equation}
and, in accordance with the  degeneracy instance of Lemma 2, we also include
\begin{equation}
\boldsymbol{k}^{(0)} = (-1,-1).
\label{eq:K0}
\end{equation}

\subsection{Theorem on Nondegenerate Ground States}

For $\boldsymbol{k}^{(\alpha)}\in\mathfrak{\mathcal{K}}$, $\alpha\in\{0,1,2,\ldots,5\}$,
we define the parameter regions $\mathcal{\mathfrak{R}}_{\alpha}\subset\mathcal{\mathfrak{R}}$
as follows: \begin{align}
\mathcal{\mathfrak{R}}_{\alpha} = 
\Big\{ & (\bm{\theta}, \lambda) \in \mathcal{\mathfrak{R}} \, \Big| \, 
 E_{\boldsymbol{k}^{(\alpha)}}(\bm{\theta}, \lambda) 
< E_{\boldsymbol{k}^{(\beta)}}(\bm{\theta}, \lambda),  
\forall \beta \in \{0, 1, 2, \ldots, 5\}, \, \beta \neq \alpha 
\Big\}, \nonumber \\
& \alpha \in \{0, 1, 2, \ldots, 5\}.
\end{align} Explicit expressions for $\mathcal{\mathfrak{R}}_{1},\mathcal{\mathfrak{R}}_{2},\mathcal{\mathfrak{R}}_{3},\mathcal{\mathfrak{R}}_{4},$
and $\mathcal{\mathfrak{R}}_{5}$ are provided in Appendix B. Note
that $\mathcal{\mathfrak{R}}_{0}$ is empty.\medskip{}
\begin{quote}
\textbf{Theorem} \newline
For any $(\bm{\theta},\lambda)\in\mathcal{\mathfrak{R}}_{\alpha_{}}\subset\mathcal{\mathfrak{R}}$, with $\alpha\in\{1,2,\ldots,5\}$, the state $\psi_{\boldsymbol{k}^{(\alpha)}}(\mathrm{\boldsymbol{x}})$ is the unique nondegenerate ground state. 
\end{quote}\medskip{}The theorem requires no explicit proof, as it follows directly from
the lemmas and the specific constructions of the parameter regions
$\mathcal{\mathfrak{R}}_{\alpha_{}}.$ \\
\\
We refer to $\mathcal{\mathfrak{R}}_{\alpha_{}}$, $\alpha\in\{1,2,\ldots,5\}$,
as the nondegenerate ground state regions.\\
\\
Graphical representations of the region $\mathcal{\mathfrak{R}}_{1}$
and the combined five regions $\mathcal{\mathfrak{R}}_{\alpha}$,
with $\alpha\in\{1,2,\ldots,5\}$, are shown in Figures 1 and 2, respectively.

\paragraph{\protect\includegraphics[width=15cm]{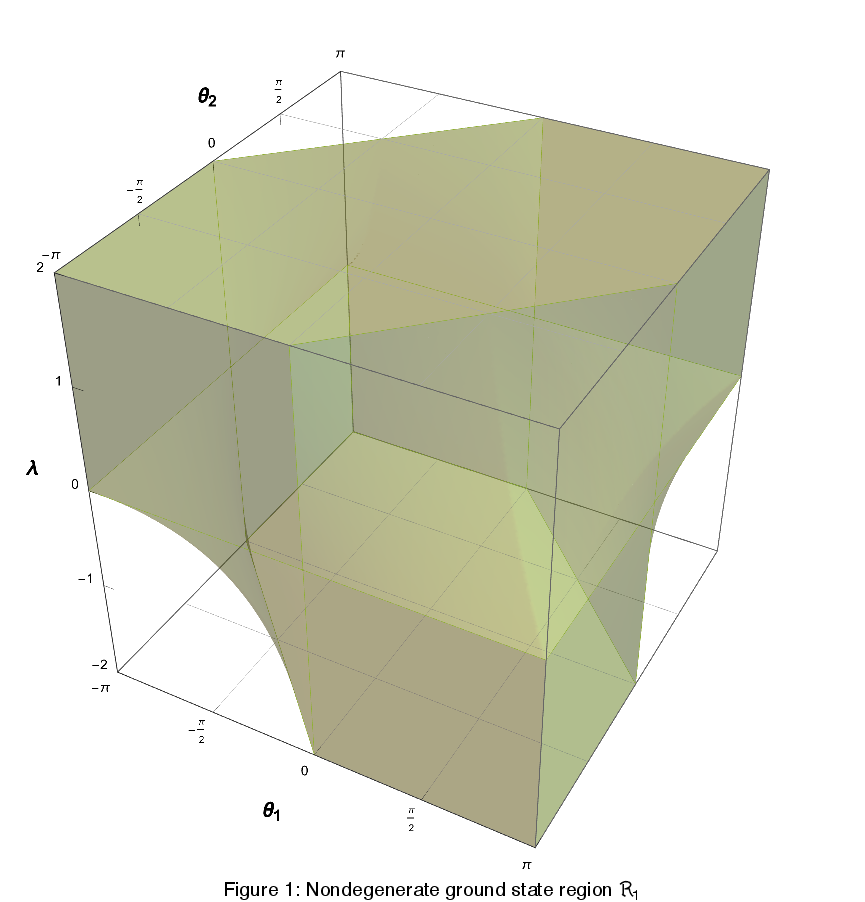}}

\paragraph{\protect\includegraphics[width=15cm]{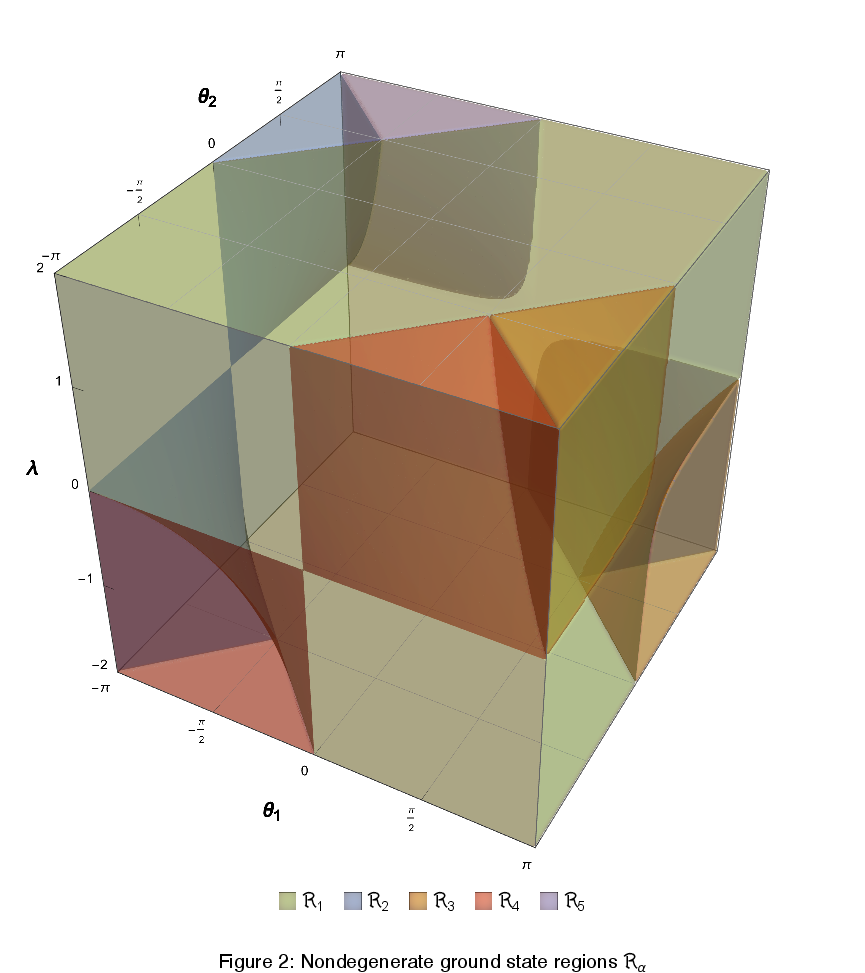}}

\subsection{Degeneracies of the Spectrum }

In the following, we systematically analyse two-, three-, and fourfold
degeneracies of the spectrum.

\subsubsection{Twofold Degenerate Ground States}

For any pair $\left(\alpha,\beta\right)$ with $\alpha\neq\beta$
and $\alpha,\beta\in\{0,1,2,\ldots,5\}$ we introduce the region $\mathcal{\mathfrak{R}}_{\alpha\beta}$
of twofold degenerate ground states for momentum vectors $\boldsymbol{k}^{(\alpha)},\boldsymbol{k}^{(\beta)}\in\mathfrak{\mathcal{K}}$
by \begin{align}
\mathcal{\mathfrak{R}}_{\alpha\beta} = 
\Big\{ & (\bm{\theta}, \lambda) \in \mathcal{\mathfrak{R}} \, \Big| \, 
E_{\boldsymbol{k}^{(\alpha)}}(\bm{\theta}, \lambda) = 
E_{\boldsymbol{k}^{(\beta)}}(\bm{\theta}, \lambda), \nonumber \\
&  E_{\boldsymbol{k}^{(\alpha)}}(\bm{\theta}, \lambda) 
< E_{\boldsymbol{k}^{(\gamma)}}(\bm{\theta}, \lambda), \quad 
\forall \gamma \in \{0,1, 2, \ldots, 5\}, \, \gamma \neq \alpha \neq \beta 
\Big\}. 
\end{align}Details about the twofold degenerate ground state regions are provided
in Appendix C. Figure 3 illustrates the regions $\,\allowbreak\mathcal{\mathfrak{R}}_{12},\,\allowbreak\mathcal{\mathfrak{R}}_{13},\,\allowbreak\mathcal{\mathfrak{R}}_{14},\textrm{and }\mathcal{\mathfrak{R}}_{15}$,
while Figures 4 and 5 depict $\,\allowbreak\mathcal{\mathfrak{R}}_{24},\,\allowbreak\mathcal{\mathfrak{R}}_{25},\,\allowbreak\mathcal{\mathfrak{R}}_{34},\textrm{and }\mathcal{\mathfrak{R}}_{35}$,
respectively, among others.\\
\\
\includegraphics[width=15cm]{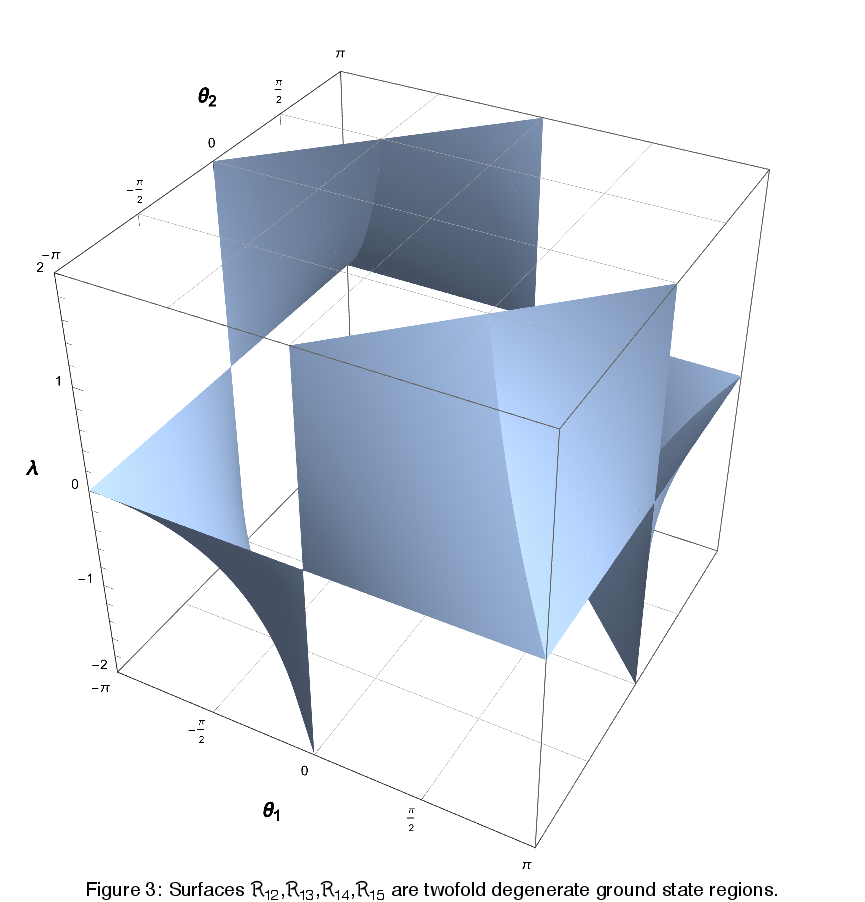}\\
\\
Finally, as a consequence of the second part of Lemma 2, a twofold degenerate ground state with degeneracy region \(\mathcal{\mathfrak{R}}_{01}\) is given by
\begin{equation}
\begin{aligned}
\mathcal{\mathfrak{R}}_{01} =\ & \Big\{ (\bm{\theta},\lambda) \in \mathcal{\mathfrak{R}} \,\big|\, E_{\boldsymbol{k}^{(0)}}(\bm{\theta},\lambda) = E_{\boldsymbol{k}^{(1)}}(\bm{\theta},\lambda) < E_{\boldsymbol{k}^{(\gamma)}}(\bm{\theta},\lambda),\\[0.2cm]
& \quad\forall \gamma \in \{2,\ldots,5\} \Big\}.
\end{aligned}
\label{eq:R01_def}
\end{equation}
This simplifies to (see Appendix C)
\begin{equation}
\mathcal{\mathfrak{R}}_{01} = \{ (\bm{\theta},\lambda) \in \mathcal{\mathfrak{R}} \,\big|\, \theta_{1} = -\pi,\, \theta_{2} = -\pi,\, \textrm{and } 0 < \lambda < 2 \},
\label{eq:R01_simplified}
\end{equation}
and is shown in Figures 4 and 5, among others.

\subsubsection{Threefold Degenerate Ground States}

For any fixed triple $\left(\alpha,\beta,\gamma\right)$ such that
$\alpha\neq\beta,\quad\alpha\neq\gamma,\quad\beta\neq\gamma$, with
$\alpha,\beta,\gamma\in\{0,1,2,\ldots,5\}$, we define the regions
$\mathcal{\mathfrak{R}}_{\boldsymbol{k}^{(\alpha)}\boldsymbol{k}^{(\beta)}\boldsymbol{k}^{(\gamma)}}$
of threefold degenerate ground states for the momentum vectors $\boldsymbol{k}^{(\alpha)},\boldsymbol{k}^{(\beta)},\boldsymbol{k}^{(\gamma)}\in\mathfrak{\mathcal{K}}$
by

\begin{align}
\mathcal{\mathfrak{R}}_{\alpha \beta \gamma} = 
\Big\{ & (\bm{\theta}, \lambda) \in \mathcal{\mathfrak{R}} \, \Big| \, 
E_{\boldsymbol{k}^{(\alpha)}}(\bm{\theta}, \lambda) = 
E_{\boldsymbol{k}^{(\beta)}}(\bm{\theta}, \lambda) = 
E_{\boldsymbol{k}^{(\gamma)}}(\bm{\theta}, \lambda) \nonumber \\
& < E_{\boldsymbol{k}^{(\delta)}}(\bm{\theta}, \lambda), \quad 
\forall \delta \in 
\{0,1, 2, \ldots, 5\}, \, \delta \neq \alpha \neq \beta \neq \gamma
\Big\}. 
\end{align}The only nonvanishing contributions are $\mathcal{\mathfrak{R}}_{124},\mathcal{\mathfrak{R}}_{125},\mathcal{\mathfrak{R}}_{134}$
and $\mathcal{\mathfrak{R}}_{135}$. Notably, no threefold degenerate
ground state involving $\boldsymbol{k}^{(0)}=\left(-1,-1\right)$
exists. Details are provided in Appendix D, with an illustration included
in Figure 4, among others. \\

\subsubsection{Fourfold Degenerate Ground States}

$\,$\\
The only instance of fourfold degeneracy $\mathcal{\mathfrak{R}}_{0125}$
arises when the vectors $\boldsymbol{k}^{(0)},\boldsymbol{k}^{(1)},\boldsymbol{k}^{(2)},$
and $\boldsymbol{k}^{(5)}$ are involved. We have\begin{equation}\mathcal{\mathfrak{R}}_{0125} =
\left\{
(\bm{\theta},\lambda) \in \mathcal{\mathfrak{R}} \, \bigg| \,
\begin{array}{l}
E_{\boldsymbol{k}^{(0)}}(\bm{\theta},\lambda) = E_{\boldsymbol{k}^{(1)}}(\bm{\theta},\lambda) = E_{\boldsymbol{k}^{(2)}}(\bm{\theta},\lambda) = E_{\boldsymbol{k}^{(5)}}(\bm{\theta},\lambda), \\
E_{\boldsymbol{k}^{(1)}}(\bm{\theta},\lambda) < 
E_{\boldsymbol{k}^{(3)}}(\bm{\theta},\lambda), \quad E_{\boldsymbol{k}^{(1)}}(\bm{\theta},\lambda) < 
E_{\boldsymbol{k}^{(4)}}(\bm{\theta},\lambda) 
\end{array}
\right\}\end{equation}which simplifies to:
\begin{equation}
\begin{aligned}\mathcal{\mathfrak{R}}_{0125} & = &  & \{(\bm{\theta},\lambda)\in\mathcal{\mathfrak{R}}\,|\,\theta_{1}=-\pi,\theta_{2}=-\pi,\lambda=0\}.\end{aligned}
\end{equation}
\\
Figure 4 provides a comprehensive view of the twofold degeneracy regions
$\mathcal{\mathfrak{R}}_{01}$, $\,\allowbreak\mathcal{\mathfrak{R}}_{12},\,\allowbreak\mathcal{\mathfrak{R}}_{13},\,\allowbreak\mathcal{\mathfrak{R}}_{14},$
and $\mathcal{\mathfrak{R}}_{15}$; it also illustrates the lines
of threefold degeneracy $\mathcal{\mathfrak{R}}_{124},\mathcal{\mathfrak{R}}_{125},\mathcal{\mathfrak{R}}_{134}$
and $\mathcal{\mathfrak{R}}_{135}$, as well as the point of fourfold
degeneracy $\mathcal{\mathfrak{R}}_{0125}$.\\
\\

\includegraphics[width=15cm]{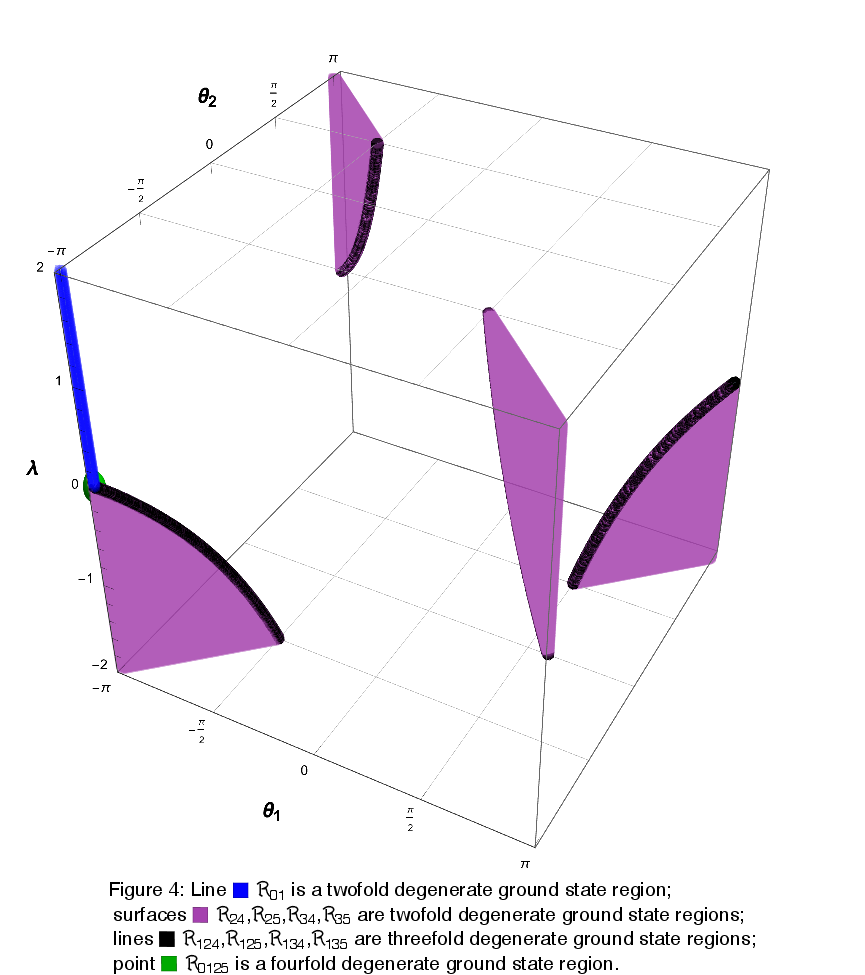}\\
\\
Complementing the representation in Figure 4, Figure 5 extends this
depiction by including all nondegenerate ground state regions $\mathcal{\mathfrak{R}}_{\alpha}$.\\
\\
\includegraphics[width=15cm]{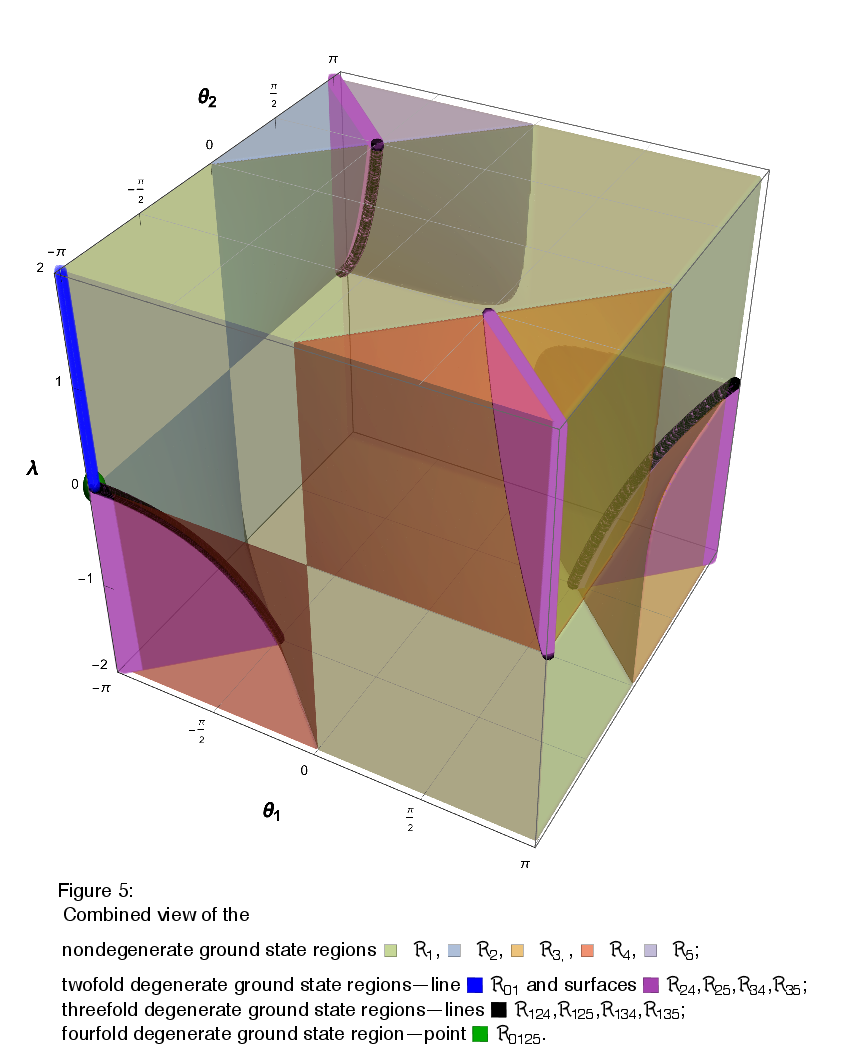}\\
\\
These scenarios of nondegenerate, two-, three-, and fourfold degenerate
ground states provide a complete classification of all possible ground-state
configurations. 

\section{Entanglement entropy}

For nondegenerate ground states, the ground-state factorises, and
consequently, the entanglement entropy vanishes. Therefore, we focus
on one of the cases of ground-state degeneracy discussed above. In
this scenario, integer-valued vectors $\boldsymbol{k}^{(\alpha)}=\left(k_{1}^{(\alpha)},k_{2}^{(\alpha)}\right)\in\mathcal{K}$
are specified, and $(\bm{\theta},\lambda)$ is restricted to the corresponding
subregion of $\mathcal{\mathfrak{R}}$. The wave functions $\psi_{\boldsymbol{k}^{(\alpha)}}(x_{1},x_{2})$
are given by
\begin{equation}
\psi_{\boldsymbol{k}^{(\alpha)}}(x_{1},x_{2})=\Lambda^{-\frac{1}{4}}e^{2\pi i\boldsymbol{k}^{(\alpha)}\boldsymbol{x}},
\end{equation}
where $\boldsymbol{x}=\left(x_{1},x_{2}\right).$The general ground-state
wave function $\Psi(x_{1},x_{2})$ is constructed as a linear superposition
of these wave functions:
\begin{equation}
\varPsi(x_{1},x_{2})=\sum_{\alpha}c_{\alpha}\psi_{\boldsymbol{k}^{(\alpha)}}(x_{1},x_{2}).
\end{equation}
where $c_{\alpha}\in\mathcal{\mathscr{\mathfrak{\mathbb{C}}}}$ are
complex coefficients, and the sum extends over the corresponding degree
of degeneracy. In ket notation we have
\begin{equation}
|\varPsi\rangle=\sum_{\alpha}c_{\alpha}|k_{1}^{(\alpha)}\rangle\otimes|k_{2}^{(\alpha)}\rangle,
\end{equation}
and $\varPsi(x_{1},x_{2})=\langle\mathrm{\boldsymbol{x}}\vert\varPsi\rangle$.\\
\\
The normalisation condition is given by
\begin{equation}
||\varPsi||^{2}=\int_{0}^{1}\int_{0}^{1}dx_{1}dx_{2}\left(det\,g\right)^{\frac{1}{2}}\varPsi^{*}(x_{1},x_{2})\varPsi(x_{1},x_{2})=\sum_{j}|c_{j}|^{2}=1.
\end{equation}
One defines the corresponding density matrix by: 
\begin{equation}
\varrho_{\varPsi}\left(x_{1},x_{2};y_{1},y_{2}\right)=\sum_{\alpha,\beta}c_{\alpha}^{*}c_{\beta}\psi_{\boldsymbol{k}^{(\alpha)}}^{*}(x_{1},x_{2})\psi_{\boldsymbol{k}^{(\beta)}}(y_{1},y_{2}).
\end{equation}
To compute the reduced density matrix, we trace out $x_{2}$:

\selectlanguage{english}%
\begin{equation}
\begin{array}{cc}
\varrho_{\varPsi}^{red}\left(x_{1},y_{1}\right) & \begin{array}{cc}
= & \int_{0}^{1}dx_{2}\varrho_{\varPsi}\left(x_{1},x_{2};y_{1},x_{2}\right),\end{array}\end{array}\label{rhored}
\end{equation}
\foreignlanguage{british}{which simplifies to 
\begin{equation}
\begin{aligned}\varrho_{\varPsi}^{red}\left(x_{1},y_{1}\right) & = & \left(det\,g\right)^{-\frac{1}{2}}\sum_{\alpha,\beta}c_{\alpha}^{*}c_{\beta}\delta_{k_{2}^{(\alpha)},k_{2}^{(\beta)}}e^{-2\pi i\left(k_{1}^{(\alpha)}x_{1}-k_{1}^{(\beta)}y_{1}\right)}\end{aligned}
.
\end{equation}
Next, to find the eigenvalues of $\varrho_{\varPsi}^{red}$ we consider
the Fredholm alternative 
\begin{equation}
\int_{0}^{1}dy_{1}\left(det\,g\right)^{\frac{1}{2}}\varrho_{\varPsi}^{red}\left(x_{1},y_{1}\right)f(y_{1})=\mu\,f(x_{1}).
\end{equation}
Using the definitions 
\begin{equation}
\varXi_{\beta\gamma}=\sum_{\alpha}c_{\alpha}^{*}c_{\gamma}\delta_{k_{1}^{(\beta)},k_{1}^{(\alpha)}}\delta_{k_{2}^{(\alpha)},k_{2}^{(\gamma)}},\,f_{\beta}=\int_{0}^{1}dx_{1}\left(det\,g\right)^{\frac{1}{2}}e^{2\pi ik_{1}^{(\beta)}x_{1}}f(x_{1}),\label{eq:Xi}
\end{equation}
where $f$ is any appropriate function on $\mathbb{S}^{1},$ this
reduces to the study of eigenvalues $\mu_{\alpha}$ of the matrix
$\varXi_{\beta\gamma}$
\begin{equation}
\sum_{\gamma}\varXi_{\beta\gamma}\,f_{\gamma}=\mu_{\alpha}\,f_{\beta}.
\end{equation}
Finally, the von Neumann entropy $\zeta_{vN}$ is obtained as:
\begin{equation}
\zeta_{vN}=-lim_{k\rightarrow1}\partial_{k}Tr\,\varXi^{k}=-\sum_{\alpha}\mu_{\alpha}log\left(\mu_{\alpha}\right).\label{eq:vNee}
\end{equation}
Before concluding this section, we investigate how the transformation
\begin{equation}
\bm{\theta}\rightarrow\bm{\theta}+2\pi\,\bm{l}\textrm{ with }l\in\mathbb{Z}^{2}
\end{equation}
affects the entanglement entropy $\zeta_{vN}$. According to (\ref{eq:shift}),
this shift changes the momenta of the ground states, so that
\begin{equation}
\boldsymbol{k}^{(\alpha)}\rightarrow\boldsymbol{k}^{(\alpha)}+\bm{l},
\end{equation}
for each $\psi_{\boldsymbol{k}^{(\alpha)}}(x_{1},x_{2})$. Importantly,
the matrix $\varXi_{\beta\gamma}$ introduced in (\ref{eq:Xi}) remains
invariant under this transformation. Consequently, the entanglement
entropy depends only on the parameter space $\mathcal{\mathfrak{R}}$,
as given in Eq. (\ref{eq:rg=00005Censuremath=00007B=00005Ctheta=00007D}).
In fact, it descends to a function on $\mathcal{\mathfrak{R}}$ that,
as our analysis shows, exhibits discontinuities.}
\selectlanguage{british}%

\subsection{Zero Entanglement Entropy for Twofold Degenerate Ground States }

Let us consider the states associated with the regions $\,\allowbreak\mathcal{\mathfrak{R}}_{12},\,\allowbreak\mathcal{\mathfrak{R}}_{13},\,\allowbreak\mathcal{\mathfrak{R}}_{14},$
and $\mathcal{\mathfrak{R}}_{15}$. In each case, the corresponding
state $|\Psi\rangle$ factorises, resulting in vanishing entanglement
entropy, see Table 2.\\
\\
For parameters $(\bm{\theta},\lambda)$ confined to these regions,
the entanglement entropy vanishes identically, highlighting regimes
of exact separability and the absence of quantum correlations across
these subsystems.

\subsection{Non-zero Entanglement Entropy for Twofold Degenerate Ground States }

Non-zero entanglement entropy emerges in regions such as $\mathcal{\mathfrak{R}}_{24},\mathcal{\mathfrak{R}}_{25},\mathcal{\mathfrak{R}}_{34},$
and $\mathcal{\mathfrak{R}}_{35}$. Consider, for example, $\mathcal{\mathfrak{R}}_{24}$,
where the corresponding ground-state wave function in ket notation
is

\begin{equation}
|\Psi\rangle=\frac{c_{1}|-1\rangle\otimes|0\rangle+c_{2}|0\rangle\otimes|-1\rangle}{\sqrt{2}},\qquad\textrm{with}\quad\sum_{j=1}^{2}|c_{j}|^{2}=1.
\end{equation}
The reduced density matrix takes the form

\begin{equation}
\varXi=\left(\begin{array}{cc}
\left|c_{1}\right|{}^{2} & 0\\
0 & \left|c_{2}\right|{}^{2}
\end{array}\right),
\end{equation}
and the von Neumann entanglement entropy then evaluates to
\begin{equation}
\zeta_{vN}=-\left|c_{1}\right|\log\left(\left|c_{1}\right|\right)-\left|c_{2}\right|\log\left(\left|c_{2}\right|\right).\label{eq:ee5.2}
\end{equation}
Similar calculations apply to the other regions, $\mathcal{\mathfrak{R}}_{24},\mathcal{\mathfrak{R}}_{34},\mathcal{\mathfrak{R}}_{35}$
and $\mathcal{\mathfrak{R}}_{01}$, yielding the same result as expressed
in Equation (\ref{eq:ee5.2}), see Table 2.

\subsection{Non-zero Entanglement Entropy for Threefold Degenerate Ground States }

Entanglement entropy becomes non-zero in regions with threefold degeneracies,
specifically in $\mathcal{\mathfrak{R}}_{124},\mathcal{\mathfrak{R}}_{125},\mathcal{\mathfrak{R}}_{134}$
and $\mathcal{\mathfrak{R}}_{135}$. To illustrate, consider $\mathcal{\mathfrak{R}}_{125}$.
The corresponding ground-state wave function in ket notation is

\begin{equation}
|\Psi\rangle=\frac{c_{1}|-1\rangle\otimes|0\rangle+c_{2}|0\rangle\otimes|1\rangle+c_{3}|0\rangle\otimes|0\rangle}{\sqrt{3}},
\end{equation}
 where $\sum_{j=1}^{3}|c_{j}|^{2}=1$. The associated matrix $\varXi$
obtains as
\begin{equation}
\varXi=\left(\begin{array}{ccc}
\left|c_{1}\right|{}^{2} & \left(c_{3}\right){}^{*}c_{2} & \left|c_{3}\right|{}^{2}\\
0 & \left|c_{2}\right|{}^{2} & \left(c_{2}\right){}^{*}c_{3}\\
\left|c_{1}\right|{}^{2} & \left(c_{3}\right){}^{*}c_{2} & \left|c_{3}\right|{}^{2}
\end{array}\right).
\end{equation}
The eigenvalues of $\varXi$ are
\begin{eqnarray}
\mu_{1} & = & 0\\
\mu_{2} & = & \frac{1}{2}\left(1-\sqrt{r_{1}^{4}+2\left(r_{3}^{2}-r_{2}^{2}\right)r_{1}^{2}+\left(r_{2}^{2}+r_{3}^{2}\right){}^{2}}\right)\\
\mu_{3} & = & \frac{1}{2}\left(1+\sqrt{r_{1}^{4}+2\left(r_{3}^{2}-r_{2}^{2}\right)r_{1}^{2}+\left(r_{2}^{2}+r_{3}^{2}\right){}^{2}}\right),
\end{eqnarray}
where $r_{i}=|c_{i}|.$ The von Neumann entanglement entropy is then
given by
\begin{equation}
\zeta_{vN}=-\mu_{2}\log\left(\mu_{2}\right)-\mu_{3}\log\left(\mu_{3}\right).
\end{equation}
Special cases include
\begin{equation}
\zeta_{vN}=\log\left(2\right)\;\textrm{for}\;r_{1}=r_{2}=\frac{1}{\sqrt{2}},\;r_{3}=0,
\end{equation}
and
\begin{equation}
\begin{aligned}\zeta_{vN}=\log\left(3\right)-\frac{1}{3}\sqrt{5}\coth{}^{-1}(\frac{3}{\sqrt{5}}) & \; & \;\textrm{for}\;r_{1}=r_{2}=r_{3}=\frac{1}{\sqrt{3}}.\end{aligned}
\end{equation}
The same von Neumann entanglement entropy result is obtained for the
regions $\mathcal{\mathfrak{R}}_{124},\mathcal{\mathfrak{R}}_{134},$
and $\mathcal{\mathfrak{R}}_{135}$, see Table 3.

\subsection{Non-zero Entanglement Entropy for Fourfold Degenerate Ground States}

Non-zero entanglement entropy arises in the case of fourfold degeneracy,
represented by the region $\mathcal{\mathfrak{R}}_{0125}$. The corresponding
ground state in ket notation is
\begin{equation}
|\Psi\rangle=\frac{1}{2}\left(c_{1}|0\rangle\otimes|0\rangle+c_{2}|-1\rangle\otimes|-1\rangle+c_{3}|-1\rangle\otimes|0\rangle+c_{4}|0\rangle\otimes|-1\rangle\right),
\end{equation}
where $\sum_{j=1}^{4}|c_{j}|^{2}=1$. This leads to the matrix
\begin{equation}
\varXi=\left(\begin{array}{cccc}
\left|c_{1}\right|{}^{2} & c_{2}\left(c_{4}\right){}^{*} & c_{3}\left(c_{1}\right){}^{*} & \left|c_{4}\right|{}^{2}\\
c_{1}\left(c_{3}\right){}^{*} & \left|c_{2}\right|{}^{2} & \left|c_{3}\right|{}^{2} & c_{4}\left(c_{2}\right){}^{*}\\
c_{1}\left(c_{3}\right){}^{*} & \left|c_{2}\right|{}^{2} & \left|c_{3}\right|{}^{2} & c_{4}\left(c_{2}\right){}^{*}\\
\left|c_{1}\right|{}^{2} & c_{2}\left(c_{4}\right){}^{*} & c_{3}\left(c_{1}\right){}^{*} & \left|c_{4}\right|{}^{2}
\end{array}\right)
\end{equation}
The eigenvalues of $\varXi$ are
\begin{eqnarray}
\mu_{1} & = & 0\\
\mu_{2} & = & 0\\
\mu_{3} & = & \frac{1}{2}\left(1-\sqrt{1-4\mu}\right)\\
\mu_{4} & = & \frac{1}{2}\left(1+\sqrt{1-4\mu}\right)
\end{eqnarray}
where $\mu=|det\left(\begin{array}{cc}
c_{1} & c_{3}\\
c_{4} & c_{2}
\end{array}\right)|.$ The von Neumann entanglement entropy is then
\begin{equation}
\zeta_{vN}=-\mu_{3}\log\left(\mu_{3}\right)-\mu_{4}\log\left(\mu_{4}\right),
\end{equation}
see Table 4. Note that $\zeta_{vN}$ vanishes if and only if $det\left(\begin{array}{cc}
c_{1} & c_{3}\\
c_{4} & c_{2}
\end{array}\right)=0.$\\
\\
It is worth noting that the emergence of non-zero entanglement entropy
in the case of fourfold degeneracy is an intrinsic feature of the
model, present at zero coupling $\lambda=0,$ under the specific conditions
$\text{\ensuremath{\theta_{1}}}\LyXZeroWidthSpace=-\pi,\text{\ensuremath{\theta_{2}}}\LyXZeroWidthSpace=-\pi.$
\\
\\
We collect all relevant momentum vectors in Table 1 and summarise
in Tables 2 to 4 the different expressions for the von Neumann entanglement
entropy corresponding to the various types of ground state degeneracies.
\begin{table}[H]
\centering
\begin{minipage}{\dimexpr0.3\textwidth+0.6\textwidth+4\tabcolsep+1.2pt\relax}
\raggedright
Table 1: Momentum Vectors
\vspace{0.3cm}

\begin{tabular}{|p{\dimexpr0.3\textwidth+0.6\textwidth+4\tabcolsep+1.2pt\relax}|}
\hline
$\boldsymbol{k}^{(0)} = (-1,-1)$ \\
$\boldsymbol{k}^{(1)} = (0,0)$ \\
$\boldsymbol{k}^{(2)} = (-1,0)$ \\
$\boldsymbol{k}^{(3)} = (1,0)$ \\
$\boldsymbol{k}^{(4)} = (0,-1)$ \\
$\boldsymbol{k}^{(5)} = (0,1)$ \\
\hline
\end{tabular}
\end{minipage}
\refstepcounter{table}
\end{table}

\begin{table}[H]
\caption{Twofold Degeneracies and von Neumann Entanglement Entropy}
\vspace{0.3cm} 
\centering
\begin{tabular}{|p{0.3\textwidth}|p{0.6\textwidth}|}
\hline
\textbf{Pairs of Momentum Vectors} & \textbf{von Neumann Entanglement Entropy} \\
\hline
\begin{tabular}[t]{@{}l@{}}
$(\boldsymbol{k}^{(1)}, \boldsymbol{k}^{(2)})$, \\
$(\boldsymbol{k}^{(1)}, \boldsymbol{k}^{(3)})$, \\
$(\boldsymbol{k}^{(1)}, \boldsymbol{k}^{(4)})$, \\
$(\boldsymbol{k}^{(1)}, \boldsymbol{k}^{(5)})$ 
\end{tabular} & $\zeta_{vN} = 0$ \\
\hline
\begin{tabular}[t]{@{}l@{}}
$(\boldsymbol{k}^{(2)}, \boldsymbol{k}^{(4)})$, \\
$(\boldsymbol{k}^{(2)}, \boldsymbol{k}^{(5)})$, \\
$(\boldsymbol{k}^{(3)}, \boldsymbol{k}^{(4)})$, \\
$(\boldsymbol{k}^{(3)}, \boldsymbol{k}^{(5)})$, \\
$(\boldsymbol{k}^{(0)}, \boldsymbol{k}^{(1)})$
\end{tabular} & 
\begin{tabular}[t]{@{}l@{}}
$\zeta_{vN} = -\left|c_{1}\right|\log(\left|c_{1}\right|) - \left|c_{2}\right|\log(\left|c_{2}\right|)$ \\[0.2cm]
$\sum_{j=1}^{2}|c_{j}|^{2}=1$
\end{tabular} \\
\hline
\end{tabular}
\end{table}

\begin{table}[H]
\caption{Threefold Degeneracies and von Neumann Entanglement Entropy}
\vspace{0.3cm} 
\centering
\begin{tabular}{|p{0.3\textwidth}|p{0.6\textwidth}|}
\hline
\textbf{Triplets of \newline Momentum Vectors} & \textbf{von Neumann Entanglement Entropy} \\
\hline
\begin{tabular}[t]{@{}l@{}}
$(\boldsymbol{k}^{(1)}, \boldsymbol{k}^{(2)}, \boldsymbol{k}^{(5)})$, \\
$(\boldsymbol{k}^{(1)}, \boldsymbol{k}^{(2)}, \boldsymbol{k}^{(4)})$, \\
$(\boldsymbol{k}^{(1)}, \boldsymbol{k}^{(3)}, \boldsymbol{k}^{(5)})$, \\
$(\boldsymbol{k}^{(1)}, \boldsymbol{k}^{(3)}, \boldsymbol{k}^{(4)})$
\end{tabular} & 
\begin{tabular}[t]{@{}l@{}}
$\zeta_{vN} = -\mu_{2}\log(\mu_{2}) - \mu_{3}\log(\mu_{3})$ \\
$\mu_{1}=0$ \\
$\mu_{2} = \frac{1}{2}\left(1-\sqrt{r_{1}^{4} + 2\left(r_{3}^{2}-r_{2}^{2}\right)r_{1}^{2} + \left(r_{2}^{2}+r_{3}^{2}\right)^{2}}\right)$ \\
$\mu_{3} = \frac{1}{2}\left(1+\sqrt{r_{1}^{4} + 2\left(r_{3}^{2}-r_{2}^{2}\right)r_{1}^{2} + \left(r_{2}^{2}+r_{3}^{2}\right)^{2}}\right)$ \\
$\text{where} \; r_{i} = |c_{i}|,\;\text{and}\;\sum_{i=1}^{3}r_{i}^{2}=1$ \\
\end{tabular} \\
\hline
\end{tabular}
\end{table}

\begin{table}[H]
\caption{Fourfold Degeneracies and von Neumann Entanglement Entropy}
\vspace{0.3cm} 
\centering
\begin{tabular}{|p{0.3\textwidth}|p{0.6\textwidth}|}
\hline
\textbf{Quartet of \newline Momentum Vectors} & \textbf{von Neumann Entanglement Entropy} \\
\hline
\small $(\boldsymbol{k}^{(0)}, \boldsymbol{k}^{(1)}, \boldsymbol{k}^{(2)}, \boldsymbol{k}^{(5)})$ & 
\begin{tabular}[t]{@{}l@{}}
$\zeta_{vN} = -\mu_{3}\log(\mu_{3}) - \mu_{4}\log(\mu_{4})$ \\
$\mu_{1}=0, \quad \mu_{2}=0$ \\
$\mu_{3} = \frac{1}{2}\left(1-\sqrt{1-4\mu}\right)$ \\
$\mu_{4} = \frac{1}{2}\left(1+\sqrt{1-4\mu}\right)$ \\
$\text{where} \; \mu = |\det\begin{pmatrix} c_{1} & c_{3} \\ c_{4} & c_{2} \end{pmatrix}|,\;\;\text{and}\;\sum_{j=1}^{4}|c_{j}|^{2}=1$ \\
\end{tabular} \\
\hline
\end{tabular}
\end{table}

\section{Degeneracies of the Ground State, Entanglement Entropy and Fixing
of the Metric}

Our investigation of quantum systems with complex spatial configurations
reveals a strong connection between nonvanishing entanglement entropy
and constraints on the parameter space. In our model---where two-,
three-, and fourfold ground state degeneracies emerge---entanglement
entropy serves as a diagnostic tool for underlying geometric and topological
structure.\\
\\
Specifically, for twofold degeneracy, the entropy is nonzero in regions
$\mathcal{\mathfrak{R}}_{24},$ $\mathcal{\mathfrak{R}}_{25},$ $\mathcal{\mathfrak{R}}_{34},$
$\mathcal{\mathfrak{R}}_{35},$ and $\mathcal{\mathfrak{R}}_{01}$,
with $\lambda$ varying within finite bounds; for threefold degeneracy,
$\lambda$ is uniquely fixed for a given $\bm{\theta}$; and for fourfold
degeneracy, both $\bm{\theta}$ and $\lambda$ are uniquely constrained,
with the metric becoming the canonical metric on the 2-torus. \\
\\
We also clarify that in regions $\mathcal{\mathfrak{R}}_{\alpha}$
where no spectral degeneracies occur, the entanglement entropy vanishes,
and that even in a nontrivial scenario with twofold degeneracy (in
regions $\mathcal{\mathfrak{R}}_{12},\mathcal{\mathfrak{R}}_{13},\mathcal{\mathfrak{R}}_{14},$
and $\mathcal{\mathfrak{R}}_{15}$), the ground state is factorised,
leading to zero entanglement entropy.\\
\\
The non-vanishing of the entanglement entropy signals the existence
of correlations between the momentum degrees of freedom of the particle---correlations
that are absent in a classical description.\\
\\
These findings contribute to the broader understanding of how coupling
constants and system parameters are intrinsically constrained---a
theme long explored via mechanisms such as renormalisation group flow
\cite{wilson,politzer,gross wilczek}, symmetry \cite{t hooft,mandelstam},
dynamical symmetry breaking \cite{nambu}, and coupling reduction
\cite{Zimmermann}.\\
\\
Our work provides a careful and original perspective by demonstrating
that quantum entanglement not only mirrors system parameters but can
actively shape them; in the case of a charged particle on a 2-torus,
our method links entanglement entropy to the underlying metric and
topology.

\part*{Acknowledgement}

We are grateful for valuable and inspiring discussions with Heide
Narnhofer and \v{C}aslav Brukner.

\appendix

\part*{Appendix}

\section{Proofs of Lemmas }

\medskip

\noindent\textbf{Lemma 1}\\
Lemma 1 is proven by demonstrating that the contrary assumption is false. Suppose there exists a  pair \(\boldsymbol{m} = (m_1, m_2) \in \mathbb{Z}^2\), with \(\mbox{\(\big||m_1| - |m_2|\big| > 1\)}\), such that
\begin{equation}
E_{0,0}(\bm{\theta},\lambda) \geq E_{m_1, m_2}(\bm{\theta},\lambda).\label{eq:lemma1new}
\end{equation}
Without loss of generality, assume that \(|m_1| > |m_2|\), otherwise, interchange \(m_1\) and \(m_2\). Since \(|m_1| - |m_2| > 1\), it follows that \(|m_1| > 1\). Considering the cases \(m_1 \gtrless 1\) and \(m_2 \lesseqgtr 0\), it suffices to examine \(m_1 > 1\), \(m_2 \geqslant 0\). From the inequality (\ref{eq:lemma1new}) we expand and rearrange terms, leading directly to:
\begin{equation}
(m_1 - m_2)\left(\theta_2 \lambda - 2\theta_1 + 2\pi (m_1 - m_2)\right) \leq (\lambda - 2)m_2\left(-\theta_1 - \theta_2 + 2\pi m_1\right).
\end{equation}
Considering that \((\bm{\theta},\lambda) \in \mathcal{\mathfrak{R}}\), a sign contradiction emerges for both \(m_2 = 0\) and \(m_2 > 0\), thus completing the proof.

\medskip

\medskip
\noindent\textbf{Lemma 2}\\
For proving the first part of Lemma 2, we proceed again by contradiction, assuming the existence of a pair \(\boldsymbol{m} = (m_1, m_2) \in \mathbb{Z}^2\), with \(\mbox{\(|m_1| - |m_2| = 0\)}\) and \(\boldsymbol{m} \neq (0,0)\), such that
\begin{equation}
E_{0,0}(\bm{\theta},\lambda) \geq E_{m_1, m_2}(\bm{\theta},\lambda). \label{eq:lemma2new}
\end{equation}
Considering the cases \(m_1 = m_2 \gtrless 0\) and \(m_1 = -m_2 \gtrless 0\), it suffices to examine \(m_1 = m_2 \gtrless 0\). We then arrive at
\begin{equation}
m_1(\lambda - 2)\left(-\theta_1 - \theta_2 + 2\pi m_1\right) \geqslant 0.
\end{equation}
Since \(\bm{\theta} \neq \bm{\bar{\theta}}=(-\pi,-\pi)\), this part of the proof concludes with a sign contradiction, similar to the arguments in proving Lemma 1. 

We now prove the second part of Lemma 2, starting from
\begin{equation}
E_{0,0}(\bm{\theta},\lambda) = E_{m_1, m_2}(\bm{\theta},\lambda). \label{eq:lemma2_2ndpart}
\end{equation}
Among the various cases, excluding \(\boldsymbol{m} = (0,0)\), the only interesting scenario arises when \(m_1 = m_2 < 0\). In this case, we obtain
\begin{equation}
m_1(\lambda - 2)\left(-\theta_1 - \theta_2 + 2\pi m_1\right) = 0.
\end{equation}
For \(-2 < \lambda < 2\), we obtain the nontrivial solution \(m_1 = m_2 = -1\) and \(\theta_1 = -\pi\), \(\theta_2 = -\pi\).

Finally, to conclude the proof of the second part of Lemma 2, we proceed again by contradiction,
\begin{equation}
E_{0,0}(\bm{\bar{\theta}}, \lambda) \geqslant E_{m_1, m_2}(\bm{\bar{\theta}}, \lambda).
\end{equation}
Considering the cases \(m_1 = m_2 \gtrless 0\) and \(m_1 = -m_2 \gtrless 0\), with \(\boldsymbol{m} \neq (0,0)\) and \(\boldsymbol{m} \neq (-1,-1)\), it suffices to examine either \(m_1 = m_2 > 0\) or \(m_1 = m_2 < -1\). We then arrive at
\begin{equation}
m_1(\lambda - 2)\left(1 + m_1\right) \geqslant 0,
\end{equation}
which unavoidably leads to a sign contradiction.

\medskip

\noindent\textbf{Lemma 3}\\
It suffices to prove only the first part of Lemma 3. As with the previous cases, Lemma 3 is proven by contradiction, assuming a pair \(\boldsymbol{m} = (m_1, m_2) \in \mathbb{Z}^2\), with \(|m_1| - |m_2| = 1\), \(m_1 > 0\), and \(\boldsymbol{m} \neq (1,0)\), for which
\begin{equation}
E_{1,0}(\bm{\theta},\lambda) \geq E_{m_1, m_2}(\bm{\theta},\lambda). \label{eq:lemma3}
\end{equation}
Considering the cases \(m_1 > 0\) and \(m_2 \gtrless 0\), it is enough to examine \(m_1 > 0\), \(m_2 > 0\). Excluding \(\boldsymbol{m} = (1,0)\), from \(m_1 - m_2 = 1\) we deduce that \(m_1 > 1\) and \(m_2 > 0\). The inequality (\ref{eq:lemma3}) directly leads to
\begin{equation}
(\lambda-2)m_2\left(-\theta_1 - \theta_2 + 2\pi (m_2+1)\right) \geq 0,
\end{equation}
thus confirming the proof once again by a sign contradiction.

\medskip

\noindent\textbf{Lemma 4}\\
Lemma 4 is proven similarly to Lemma 3. It again suffices to prove only the first part of the lemma. As  previously, Lemma 4 is proven by contradiction, assuming a pair \(\boldsymbol{m} = (m_1, m_2) \in \mathbb{Z}^2\), with \(|m_1| - |m_2| = -1\), \(m_2 > 0\), and \(\boldsymbol{m} \neq (0,1)\), for which
\begin{equation}
E_{0,1}(\bm{\theta},\lambda) \geq E_{m_1, m_2}(\bm{\theta},\lambda).  \label{eq:lemma4new} 
\end{equation}
Considering the cases \(m_2 > 0\) and \(m_1 \gtrless 0\), it is enough to examine \(m_1 > 0\), \(m_2 > 0\). Excluding \(\boldsymbol{m} = (0,1)\), from \(m_1 - m_2 = -1\) we deduce that \(m_1 > 0\) and \(m_2 > 1\). The inequality (\ref{eq:lemma4new}) directly leads to
\begin{equation}
(\lambda-2)m_1\left(-\theta_1 - \theta_2 + 2\pi (m_1+1)\right) \geq 0.
\end{equation}
This completes the proof, once again by a sign contradiction, as in the previous cases.

\section{Nondegenerate Ground States}

We present explicit expressions for the parameter regions $\mathcal{\mathfrak{R}}_{\alpha}$.
The task of reducing and logically expanding all parameter restrictions,
supported by Mathematica, resulted in intricate expressions, particularly
for $\mathcal{\mathfrak{R}}_{1}$. We now provide a more concise representation
of these results:

\begin{equation}
\begin{aligned}\mathcal{\mathfrak{R}}_{1} & = & \{(\bm{\theta},\lambda)\in\mathcal{\mathfrak{R}}\,| & \,\theta_{1}\lambda<2\left(\theta_{2}+\pi\right),\theta_{2}\lambda<2\left(\theta_{1}+\pi\right),\\
 &  &  & \,2\theta_{1}<\theta_{2}\lambda+2\pi,2\theta_{2}<\theta_{1}\lambda+2\pi,\\
 &  &  & \,2\pi+\theta_{1}+\theta_{2}>0\},
\end{aligned}
\end{equation}

\begin{equation}
\begin{aligned}\mathcal{\mathfrak{R}}_{2} & = &  & \{(\bm{\theta},\lambda)\in\mathcal{\mathfrak{R}}\,|\,\theta_{1}-\theta_{2}<0,\theta_{1}+\theta_{2}<0,\\
 &  &  & 2\left(\theta_{1}+\pi\right)<\theta_{2}\lambda,\left(\theta_{1}+2\pi\right)\lambda<2\left(\theta_{2}+\pi\right)\},
\end{aligned}
\end{equation}

\begin{equation}
\begin{aligned}\mathcal{\mathfrak{R}}_{3} & = &  & \{(\bm{\theta},\lambda)\in\mathcal{\mathfrak{R}}\,|\,\theta_{1}+\theta_{2}>0,\theta_{1}-\theta_{2}>0,\\
 &  &  & \theta_{2}\lambda+2\pi<2\theta_{1},\left(\theta_{1}+2\theta_{2}+2\pi\right)\lambda<2\left(2\theta_{1}+\theta_{2}+\pi\right)\},
\end{aligned}
\end{equation}

\begin{equation}
\begin{aligned}\mathcal{\mathfrak{R}}_{4} & = &  & \{(\bm{\theta},\lambda)\in\mathcal{\mathfrak{R}}\,|\,\theta_{1}-\theta_{2}>0,\theta_{1}+\theta_{2}<0,\\
 &  &  & 2\left(\theta_{2}+\pi\right)<\theta_{1}\lambda,\left(\theta_{2}+2\pi\right)\lambda<2\left(\theta_{1}+\pi\right)\},
\end{aligned}
\end{equation}

\begin{equation}
\begin{aligned}\mathcal{\mathfrak{R}}_{5} & = &  & \{(\bm{\theta},\lambda)\in\mathcal{\mathfrak{R}}\,|\,\theta_{1}+\theta_{2}>0,\theta_{1}-\theta_{2}<0,\\
 &  &  & \theta_{1}\lambda+2\pi<2\theta_{2},\left(2\left(\theta_{1}+\pi\right)+\theta_{2}\right)\lambda<2\left(\theta_{1}+2\theta_{2}+\pi\right)\}.
\end{aligned}
\end{equation}

\section{Twofold degenerate Ground States}

Explicit expressions are given for the parameter regions $\mathcal{\mathfrak{R}}_{\alpha\beta}$.
The non-vanishing results, presented concisely, are as follows:

\begin{equation}
\begin{aligned}\mathcal{\mathfrak{R}}_{12} & = &  & \{(\bm{\theta},\lambda)\in\mathcal{\mathfrak{R}}\,|\,2\left(\theta_{1}+\pi\right)=\theta_{2}\lambda,\theta_{1}\lambda<2\left(\theta_{2}+\pi\right),2\theta_{2}<\theta_{1}\lambda+2\pi\},\end{aligned}
\end{equation}
\begin{equation}
\begin{aligned}\mathcal{\mathfrak{R}}_{13} & = &  & \{(\bm{\theta},\lambda)\in\mathcal{\mathfrak{R}}\,|\,\theta_{2}\lambda+2\pi=2\theta_{1},2\theta_{2}<\theta_{1}\lambda+2\pi,\theta_{1}\lambda<2\left(\theta_{2}+\pi\right)\},\end{aligned}
\end{equation}
\begin{equation}
\begin{aligned}\mathcal{\mathfrak{R}}_{14} & = &  & \{(\bm{\theta},\lambda)\in\mathcal{\mathfrak{R}}\,|\,2\left(\theta_{2}+\pi\right)=\theta_{1}\lambda,2\theta_{1}<\theta_{2}\lambda+2\pi,\theta_{2}\lambda<2\left(\theta_{1}+\pi\right)\},\end{aligned}
\end{equation}

\begin{equation}
\begin{aligned}\mathcal{\mathfrak{R}}_{15} & = &  & \{(\bm{\theta},\lambda)\in\mathcal{\mathfrak{R}}\,|\,\theta_{1}\lambda+2\pi=2\theta_{2},\theta_{2}\lambda<2\left(\theta_{1}+\pi\right),2\theta_{1}<\theta_{2}\lambda+2\pi\},\end{aligned}
\end{equation}

\begin{equation}
\begin{aligned}\mathcal{\mathfrak{R}}_{24} & = &  & \{(\bm{\theta},\lambda)\in\mathcal{\mathfrak{R}}\,|\,\theta_{1}=\theta_{2},2\left(\theta_{1}+\pi\right)<\theta_{2}\lambda\},\end{aligned}
\end{equation}

\begin{equation}
\begin{aligned}\mathcal{\mathfrak{R}}_{25} & = &  & \{(\bm{\theta},\lambda)\in\mathcal{\mathfrak{R}}\,|\,\theta_{1}=-\theta_{2},2\left(\theta_{1}+\pi\right)<\theta_{2}\lambda\},\end{aligned}
\end{equation}

\begin{equation}
\begin{aligned}\mathcal{\mathfrak{R}}_{34} & = &  & \{(\bm{\theta},\lambda)\in\mathcal{\mathfrak{R}}\,|\,\theta_{1}=-\theta_{2},\theta_{2}\lambda+2\pi<2\theta_{1}\},\end{aligned}
\end{equation}

\begin{equation}
\begin{aligned}\mathcal{\mathfrak{R}}_{35} & = &  & \{(\bm{\theta},\lambda)\in\mathcal{\mathfrak{R}}\,|\,\theta_{1}=\theta_{2},\theta_{2}\lambda+2\pi<2\theta_{1}\},\end{aligned}
\end{equation}
and finally 
\begin{equation}
\mathcal{\mathfrak{R}}_{01}=\{(\bm{\theta},\lambda)\in\mathcal{\mathfrak{R}}\,|\,\theta_{1}=-\pi,\theta_{2}=-\pi,0<\lambda<2\}.
\end{equation}

\section{Threefold degenerate Ground States}

Here explicit expressions are given for $\mathcal{\mathfrak{R}}_{\alpha\beta\gamma}$.
Non vanishing results are $\mathcal{\mathfrak{R}}_{124},\mathcal{\mathfrak{R}}_{125},\mathcal{\mathfrak{R}}_{134}$
$\mathcal{\mathfrak{R}}_{135}$:

\begin{equation}
\begin{aligned}\mathcal{\mathfrak{R}}_{124} & = &  & \{(\bm{\theta},\lambda)\in\mathcal{\mathfrak{R}}\,|\,-\pi\leq\theta_{1}<-\frac{\pi}{2},\theta_{2}=\theta_{1},\lambda=\frac{2\theta_{1}+2\pi}{\theta_{2}}\},\end{aligned}
\end{equation}

\begin{equation}
\begin{aligned}\mathcal{\mathfrak{R}}_{125} & = &  & \{(\bm{\theta},\lambda)\in\mathcal{\mathfrak{R}}\,|\,-\pi<\theta_{1}<-\frac{\pi}{2},\theta_{2}=-\theta_{1},\lambda=\frac{2\theta_{1}+2\pi}{\theta_{2}}\},\end{aligned}
\end{equation}
\begin{equation}
\begin{aligned}\mathcal{\mathfrak{R}}_{134} & = &  & \{(\bm{\theta},\lambda)\in\mathcal{\mathfrak{R}}\,|\,\frac{\pi}{2}<\theta_{1}<\pi,\theta_{2}=-\theta_{1},\lambda=\frac{2\theta_{1}-2\pi}{\theta_{2}}\},\end{aligned}
\end{equation}
\begin{equation}
\begin{aligned}\mathcal{\mathfrak{R}}_{135} & = &  & \{(\bm{\theta},\lambda)\in\mathcal{\mathfrak{R}}\,|\,\frac{\pi}{2}<\theta_{1}<\pi,\theta_{2}=\theta_{1},\lambda=\frac{2\theta_{1}-2\pi}{\theta_{2}}\}.\end{aligned}
\end{equation}

\end{document}